\documentclass[11pt,a4paper]{article}
\usepackage{amssymb}
\usepackage{graphicx}
\usepackage{slashed}
\usepackage{amsmath}
\usepackage{verbatim}
\usepackage{tabulary}
\usepackage{booktabs}
\usepackage{jheppub}

\newcommand{\eql}[2]{\begin{equation}\label{#1}{\begin{split}#2\end{split}}\end{equation}}

\newcommand{\eq}[2][ ]{\begin{equation}\label{#1}{\begin{split}#2\end{split}}\end{equation}}

\newcommand{\Tb}{\overline T}

\title{The $T\Tb$ deformation at large central charge}

\author{
Ofer Aharony and Talya Vaknin
\\
\it{Department of Particle Physics and Astrophysics,\\
Weizmann Institute of Science, Rehovot 7610001, Israel}
}
\emailAdd{Ofer.Aharony@weizmann.ac.il}
\emailAdd{Talya.Vaknin@weizmann.ac.il}

\abstract{
We study Zamolodchikov's $T\Tb$ deformation of two dimensional quantum field theories in a 't Hooft-like limit, in which we scale the number of degrees of freedom $c$ to infinity and the deformation parameter $t$ to zero, keeping their product $t\cdot c$ fixed (more precisely, we keep energies and distances fixed in units of $t\cdot c$). In this limit the Hagedorn temperature remains fixed, but other non-local aspects of the theory disappear. We show that in this limit correlation functions may be computed exactly, and they are local in space and polynomials in $t$. We compute explicitly the deformed three-point functions of the energy-momentum tensor for a $T\Tb$-deformed conformal field theory.
}

\begin{document}
\maketitle
\section{Introduction}

In  \cite{Zamolodchikov:2004ce}, Zamolodchikov introduced a specific operator in two dimensional quantum field theories called $T \Tb$ (described explicitly in the next section) which has special properties, and in \cite{Cavaglia:2016oda,Smirnov:2016lqw} irrelevant deformations by this operator were studied. 
The spectrum of the deformed theory compactified on a spatial circle may be computed exactly as a function of the deformation parameter $t$ (which multiplies the $T\Tb$ operator in the action), in terms of the spectrum of the original, undeformed, theory. 
Smirnov and Zamolodchikov \cite{Smirnov:2016lqw} showed that the $S$-matrix deforms under the $T\Tb$ deformation by an overall phase which is a CDD factor \cite{Mussardo:1999aj}; in particular if the original theory is integrable then it remains so also after the deformation. In \cite{Cavaglia:2016oda} it was shown that the deformation of a theory of 24 free scalars gives the Nambu-Goto action. Other studies of these theories involve non-trivial backgrounds \cite{Cardy:2015xaa} and hydrodynamics \cite{Bernard:2015bba}. A closely related Lorentz-breaking deformation was studied in \cite{Guica:2017lia}.

A priori the deformed theories are only defined perturbatively in the deformation parameter $t$, which has units of length squared. So one has a perturbative expansion in $t*E^2$ where $E$ is some typical energy scale, which breaks down at high energies. This raises the question of what the UV completion of these theories is. One of the interesting facts about these theories is that their density of states grows exponentially, namely they have Hagedorn behavior. More precisely, this is true when the sign of the deformation parameter $t$ is negative (in standard conventions used in the literature), which we assume here (though all of our computations are valid for both signs of $t$). For this sign the energy spectrum on a circle is real for large radii, but the ground state energy becomes complex for small radii, corresponding to the Hagedorn instability. If we initially start from a conformal field theory of central charge $c$, then this happens at a radius $R$ obeying $tc = -\frac{3R^2}{2\pi}$. This implies that the deformed theories have a maximal temperature for which they are well-defined, and that they may not be local theories at high energies. A specific interesting suggestion is that \cite{Dubovsky:2017cnj} the deformation is equivalent to coupling the theory to Jackiw-Teitelboim \cite{Jackiw:1984je,Teitelboim:1983ux} gravity, such that at high energies it is a gravitational theory with no local degrees of freedom (see also \cite{Cottrell:2018skz}). For positive $t$ there are states with complex energies at all radii and the interpretation of these theories is not clear.

A recent study \cite{Cardy:2018sdv} showed that for some purposes the $T\Tb$ deformation has no local effect, since it can be rewritten as a boundary term. This allowed the derivation of differential equations for the $t$-dependence of the partition functions of $T\Tb$ deformed theories on various manifolds. However, for the purpose of computing local correlation functions, deformation terms proportional to the equations of motion, that were ignored in \cite{Cardy:2018sdv}, must be kept. We show here that local correlation functions in these theories do have a non-trivial dependence on $t$, so that the deformation can be locally felt\footnote{Similar computations were recently performed in \cite{Kraus:2018xrn}.}. In particular this is true for correlators of the energy-momentum tensor, indicating that the theory is sensitive to local changes in the background metric.

In this paper we analyze a specific limit in which correlation functions of $T\Tb$-deformed theories can be analyzed exactly. This is the large $c$ limit -- a limit of a large number of degrees of freedom, analogous to the large $N$ limit of non-Abelian gauge theories\footnote{For conformal field theories this is the limit of a large central charge, but we use the same notation also for more general quantum field theories with the property that all connected correlation functions of the energy-momentum tensor scale as some constant $c$.}. We show that in this limit, when keeping fixed $tc$, and working to leading order in $1/c$, the correlation functions are actually polynomials in $tc$, and their large $t$ (short distance) behavior is controlled just by the 2-point functions of the undeformed theory. Note that in this study we are analyzing energies that are fixed in units of $1/\sqrt{|t|c}$; they can be large in these units, but they are still much smaller than the scale $1/\sqrt{|t|}$. For these energies the correlation functions look like those of a standard local (but not scale-invariant) field theory. Some non-local features of these theories, like their Hagedorn behavior, arise at the scale $1/\sqrt{|t|c}$, while others, like the change in the S-matrix, appear at the scale $1/\sqrt{|t|}$. We do not have anything to say here about the behavior at the scale $1/\sqrt{|t|}$.

At large $c$ it is also natural to study this theory using gauge/gravity duality. The same scaling of $t$ with $c$ is required there for the deformation to be seen in classical gravity. At leading order it is a ``double-trace deformation'', which corresponds to changing the boundary conditions for the graviton in the dual gravitational theory. By construction, the holographic picture (see \cite{Shyam:2017znq,Kraus:2018xrn,Cottrell:2018skz}) will agree with our computations at the leading order in $1/c$. 
It was conjectured in \cite{McGough:2016lol} (for the positive sign of $t$) that one can also go beyond this leading order and describe the theory by putting a radial cutoff on the dual three-dimensional gravity theory, which couples the field theory to two dimensional gravity. We will not discuss the holographic picture here.

In this paper we only study correlation functions in infinite flat space; it would be interesting to generalize our results to compact and/or curved spaces, and to study what can be said there. It would also be interesting to consider the $1/c$ corrections to our results, and to understand the transition between the behavior at energy scales of order $1/\sqrt{|t|c}$ to the behavior at energies of order $1/\sqrt{|t|}$.

Note that there is a closely related deformation analyzed in \cite{Giveon:2017nie,Giveon:2017myj,Asrat:2017tzd,Giribet:2017imm}, for which the UV completion is given by a ``little string theory". This deformation can also be studied in the large $c$ limit in which it has an interesting gravitational dual. In ``little string theories'' there can also be a separation of scales between the Hagedorn scale and the non-locality scale, which is similar to what we find here. However, in this paper we do not discuss this alternative deformation but rather the original one of \cite{Zamolodchikov:2004ce,Cavaglia:2016oda,Smirnov:2016lqw}.

\section{The deformation and its effect on correlation functions}

In this section we will show how to compute correlation functions of energy momentum tensors in a quantum field theory (QFT) which is deformed by a ``$T\Tb$ deformation'', first introduced in \cite{Zamolodchikov:2004ce,Cavaglia:2016oda,Smirnov:2016lqw}. This is a short-hand notation for a family of field theories parameterized by a coupling contant $t$, which are perturbatively defined such that the change in the action when infinitesimally increasing $t$ by $\delta t$ is proportional to $\det(T)$,
\eql{tt}{S(t+\delta t) - S(t) = -{8\delta t\int \epsilon_{ik}\epsilon_{jl}T_{ij}(x)T_{kl}(x)d^2x}.}
Here $T$ is the energy-momentum tensor of the theory with parameter $t$. It depends non-trivially on $t$, so that we cannot write the deformation in a simple way using the original energy-momentum tensor (we will leave the dependence of $T$ on $t$ implicit for simplicity). We assume that for every $t$ the theory has a conserved symmetric energy-momentum tensor which is a local operator; perturbatively in $t$ this is definitely true. As shown in  \cite{Zamolodchikov:2004ce}, this particular combination of energy-momentum tensors has no singularity when the operators are brought together, so that it is a well-defined operator.

We will work in infinite flat space in Euclidean signature, where the infinitesimal deformation corresponds to inserting in the path integral
\eq{e^{8\delta t\int\epsilon_{ik}\epsilon_{jl}T_{ij}(x)T_{kl}(x)d^2x}.}
Latin letters will denote flat-space coordinates, without distinguishing upper and lower indices.

Computing correlation functions with this deformation to all orders in perturbation theory in $t$ is complicated in general. One complication is that it is not clear how general local operators ${\cal O}(x)$ should depend on $t$, since operators can mix in a complicated way under the deformation. For this reason, we will focus in this paper on correlation functions of $T$ itself, since its variation is fixed by \eqref{tt} (consistent with its conservation and symmetry). Consider an $n$-point correlation function of energy-momentum tensors,
$\langle T_{m_1n_1}(x_1)...T_{m_nn_n}(x_n)\rangle$. The definition of the theory above implies that it obeys the differential equation\footnote{We assume here for simplicity that the one-point function of $\det(T)(x)$, related \cite{Zamolodchikov:2004ce} to the square of the one-point function of $T^m_m(x)$, vanishes. On the plane this can always be achieved by a constant shift (possibly depending on $t$) in $T^m_m(x)$.}
\eql{corrchange}{\frac{d}{dt}\left\langle T_{m_1n_1}(x_1)\cdots T_{m_nn_n}(x_n)ֿ\right\rangle=&8\int d^2x\left\langle T_{m_1n_1}(x_1) \cdots T_{m_nn_n}(x_n)ֿ \epsilon_{ik} \epsilon_{jl} T_{ij}(x) T_{kl}(x)
\right\rangle\\&+\sum_{i=1}^n\left\langle T_{m_1n_1}(x_1)\cdots \frac{dT_{m_in_i}(x_i)}{dt}\cdots T_{m_nn_n}(x_n)\right\rangle ,}
with the initial condition at $t=0$ given by the correlation function in the original, undeformed theory. 
At the leading non-trivial order in $t$,
a similar computation was performed in \cite{Kraus:2018xrn} for a deformed conformal field theory. However, going to higher orders is difficult, since at $k$'th order in perturbation theory, the deformation in the $n$-point function is related to $(n+2k)$-point functions of the original theory. 

We will show in this section that the computation can be simplified by using Ward identities, and then in the next section we will show that it simplifies even further in the limit of a large number of degrees of freedom.
We follow the notations of Cardy \cite{Cardy:2018sdv} throughout this section. 

We begin by performing a Hubbard-Stratonovich transformation with an auxiliary symmetric tensor field $h_{ij}$ to rewrite
\eql{defh}{e^{8\delta t\int\epsilon_{ik}\epsilon_{jl}T_{ij}T_{kl}d^2x}\propto\int[dh]e^{-(1/32\delta t)\int\epsilon_{ik}\epsilon_{jl}h_{ij}h_{kl}d^2x+\int h_{ij}T_{ij}d^2x}.}
A general symmetric tensor field may be decomposed locally as
\eql{hdecomp}{h_{ij}=\partial_i\alpha_j+\partial_j\alpha_i+\delta_{ij}\Phi,}
where $\alpha_i(x)$ is a vector field and $\Phi(x)$ a scalar field; globally there may be obstructions to this, but in this paper we only work in flat space and $h_{ij}$ is infinitesimal so we can ignore them. Since $h_{ij}$ in \eqref{defh} is infinitesimal and couples to the energy-momentum tensor, we can treat it as an infinitesimal change of the metric of the space that the field theory lives on, and we can view \eqref{hdecomp} as its decomposition into an infinitesimal diffeomorphism $\alpha_i(x)$, and 
an infinitesimal deformation $e^\Phi$  of the conformal factor.

When writing the deformation in terms of this decomposition, the quadratic terms in $h_{ij}$, multiplying $(1/32\delta t)$ in \eqref{defh}, have the following contributions:
\eql{alphas}{2\epsilon_{ik}\epsilon_{jl}(\partial_i\alpha_j)(\partial_l\alpha_k)+2\epsilon_{ik}\epsilon_{jl}(\partial_i\alpha_j)(\partial_k\alpha_l),}
and
\eq{4\epsilon_{ik}\epsilon_{jl}(\partial_i\alpha_j)(\delta_{kl}\Phi)+\epsilon_{ik}\epsilon_{jl}(\delta_{ij}\Phi)(\delta_{kl}\Phi)=4(\partial_k\alpha_k)\Phi+2\Phi^2.}
The second term in \eqref{alphas} is a total derivative, so it can be turned into a boundary term, and we will drop it since we are interested in flat space and the correlation functions we are interested in decay fast enough such that it will not contribute. Unlike in \cite{Cardy:2018sdv}, we will keep the terms in the action that are proportional to the equations of motion, since they will be important (without these terms the full deformation is a total derivative \cite{Cardy:2018sdv}). The path integral over $h$ now takes the form
\eq{\int[d\Phi][d\alpha_i]e^{-(1/32\delta t)\int\left(2\epsilon_{ik}\epsilon_{jl}(\partial_i\alpha_j)(\partial_l\alpha_k)+4(\partial_k\alpha_k)\Phi+2\Phi^2\right) d^2x+\int\left(2\partial_i\alpha_jT_{ij}+\Phi T_{ii}\right)d^2x}.}
Integrating out $\Phi$ yields an action quadratic in $\alpha$,
\eq{\int[d\alpha_i]e^{-(1/16\delta t)\int\alpha_j\partial_i^2\alpha_jd^2x+\int\left(\alpha_j\partial_j T_{ii}-2\alpha_j\partial_i T_{ij}+4\delta t(T_{ii})^2\right)d^2x}.}
At this point we will move to momentum space and perform the integration over $\alpha_i$ to get
\eq{e^{-16\delta t \int d^2 p \frac{1}{p^2}\left(p_kT_{jk}(p) p_{k'}T_{jk'}(-p) -p_kT_{jk}(p)p_jT_{ll}(-p)\right)}.}
This form of the deformation is non-local, but it will be convenient for our computations.
Although equivalent to \eqref{tt}, this form of the action deformation exhibits the fact that one can use Ward identities to simplify the effect of the deformation, since $p_k T_{jk}(p)$ vanishes by the equations of motion, and its correlation functions are given by a sum of contact terms with other operators :
\eql{ward}{p_k \langle T_{jk}(p) {\cal O}_1(p_1)\cdots {\cal O}_n(p_n)  \rangle = \sum_{i=1}^n \langle {\cal O}_1(p_1)\cdots (\delta_j {\cal O}_i)(p_i+p) \cdots O_n(p_n) \rangle,}
where $\delta_j {\cal O}_i$ is the variation of the operator ${\cal O}_i$ under an infinitesimal translation in the $j$ direction (for scalars this is just the derivative in the $j$ direction, otherwise there are additional terms).
We can use this to simplify the computation of $n$-point correlation functions to all orders of perturbation theory.

Going back to our $n$-point correlation function $\langle T_{m_1 n_1}(x_1)\cdots T_{m_n n_n}(x_n)\rangle$, we can now rewrite our differential equation \eqref{corrchange} for the momentum-space correlation function (dropping the momentum-conservation delta function) as 
\eql{dif}{\frac{d}{dt}&\langle T_{m_1n_1}(p_1)\cdots T_{m_nn_n}(p_n)\rangle\\=&16\int d^2w\frac{1}{w^2}\langle T_{m_1n_1}(p_1)\cdots T_{m_nn_n}(p_n) w_kT_{jk}(w) \left(w_{k'}T_{jk'}(-w)-w_jT_{ll}(-w)\right)\rangle\\&+\sum_{i=1}^n\left\langle T_{m_1n_1}(p_1)
\cdots \frac{dT_{m_in_i}(p_i)}{dt}\cdots T_{m_nn_n}(p_n)\right\rangle .}
Using the Ward identity we can rewrite the first line on the right-hand side as the sum of an $n$-point function of energy-momentum tensors and an $(n+1)$-point function of energy-momentum tensors (using the fact that the variation of the energy-momentum tensor under translations is the sum of derivatives of energy-momentum tensors).
This allows us to write the variation in the $n$-point correlator in terms of $(n+1)$-point correlators at most.  Similar simplifications arise for correlation functions of other operators, but as mentioned above, for them there is no preferred choice of their $t$-derivative, so we will not analyze them here.



\section{Correlation functions in the large $c$ limit}

There is a special limit in which we can explicitly compute the correlation functions discussed in the previous section, which is the limit of a large number of degrees of freedom. For a conformal field theory (CFT) this is the large $c$ limit, where $c$ is the central charge, but there is a similar limit for general quantum field theories (which we will still call the large $c$ limit). In the large $c$ limit correlation functions of $T_{mn}$ factorize. The connected contribution to an $n$-point function of energy-momentum tensors is proportional to $c$ at large $c$, so that when we compute a general correlation function and look at the contribution to it which is a product of $k$ connected components, then this will scale as $c^k$, and the correlation function will be dominated by the contribution with the largest value of $k$. Other contributions will be suppressed by powers of $c$. In other words, for even $n$ at leading order in $c$
\eq{\langle T_{m_1n_1}(x_1)\cdots T_{m_nn_n}(x_n)\rangle = & \langle T_{m_1n_1}(x_1)T_{m_2 n_2}(x_2) \rangle \cdots \langle T_{m_{n-1} n_{n-1}}(x_{n-1}) T_{m_nn_n}(x_n)\rangle +  \\ & \qquad {\rm other\ pairings}.}
The same scaling of correlators with $c$ occurs for a more general class of operators called ``single-trace operators''. Their products (after removing singularities) are called ``multi-trace operators''. The $T\Tb$ deformation operator is an example of a double-trace operator.

Suppose we now consider the variation \eqref{corrchange} of a connected correlation function of $n$ energy-momentum tensors. In the first term on the right-hand side, the leading contribution will come from products of two connected correlation functions, with the first one including $T_{ij}(x)$ and some of the original $n$ operators, and the other including $T_{kl}(x)$ and the other original operators. This will scale as one extra power of $c$ compared to the connected contribution to this first term, and also compared to the correlation function on the left-hand side. Thus, if we want to have a finite large $c$ limit, we need to scale $t$ as $1/c$, or in other words take an 't Hooft-like limit, in which we take $c$ to infinity, keeping $tc$ finite. In this limit only the disconnected contributions to the first term on the right-hand side in \eqref{corrchange} will survive. Since $t$ has dimensions of length squared, this means that we will be computing correlation functions at distances of order $\sqrt{|t|c}$, or at energies of order $1/\sqrt{|t|c}$, in the deformed theory. We will assume this scaling in the rest of this paper \footnote{With this scaling the full action of the deformed theory scales as $c$, suggesting that a saddle point approximation may be used for computing correlation functions and other objects, but we will not use this here.}.

The change in the Lagrangian at first order in $t$ is a double-trace operator, and thus the change in $T_{mn}$ at first order is in general a combination of a double-trace and a single-trace operator. Plugging this back into the action, one can show that the terms of order $t^k$ in the action are products of $(k+1)$ single-trace operators or less, and thus the same is true also for the energy-momentum tensor. In other words,
\eq{S = \sum_{k=0}^{\infty} t^k S^{(k)}, \qquad T_{mn} = \sum_{k=0}^{\infty} t^k T_{mn}^{(k)},}
where $S^{(k)}$ and $T_{mn}^{(k)}$ include products of up to $(k+1)$ single-trace operators.

In general we could have a one-point function $\langle T_{mn}(x) \rangle = f(t) \delta_{mn}$, but in flat space we can always subtract this away without affecting the conservation equation, so we will assume that $\langle T_{mn}(x) \rangle = 0$. For two-point functions we then obtain from \eqref{dif}
\eql{diftwo}{\frac{d}{dt}&\langle T_{m_1n_1}(p_1)T_{m_2n_2}(p_2)\rangle\\=&16\int d^2w\frac{1}{w^2}\langle T_{m_1n_1}(p_1) w_kT_{jk}(w) \rangle\left \langle T_{m_2 n_2}(p_2)  \left(w_{k'}T_{jk'}(-w)-w_jT_{ll}(-w)\right)\right\rangle + (p_1 \leftrightarrow p_2) \\&+ \left\langle \frac{dT_{m_1n_1}(p_1)}{dt}
T_{m_2n_2}(p_2)\right\rangle +  \left\langle T_{m_1n_1}(p_1)
\frac{dT_{m_2n_2}(p_2)}{dt} \right\rangle  .}
The first line on the right-hand side vanishes, since using the Ward identity the first correlation function on this line is a one-point function which vanishes. The term with $dT/dt$ needs to be of order $c^2$ to contribute in our large $c$ limit, but it cannot factorize into a product of two correlators, so its contribution is suppressed at least by a factor of $c$ in the limit that we are interested in. Thus, in the large $c$ limit we obtain  that
\eq{\langle T_{m_1n_1}(p_1)T_{m_2n_2}(p_2)\rangle = \langle T_{m_1n_1}(p_1)T_{m_2n_2}(p_2)\rangle|_{t=0},}
and two-point functions are independent of $t$ \footnote{This result was previously obtained by M. Porrati (D. Kutasov, private communication).}.

Considering next $3$-point functions, and using the Ward identities, we have
\eql{difthree}{\frac{d}{dt}&\langle T_{m_1n_1}(p_1)T_{m_2n_2}(p_2) T_{m_3n_3}(p_3)\rangle\\=&-16\int d^2w\frac{1}{w^2}\langle T_{m_1n_1}(p_1) T_{m_2 n_2}(p_2) w_kT_{jk}(w) \rangle \langle T_{m_3 n_3}(p_3)  w_jT_{ll}(-w)\rangle + (2\text{ permutations}) \\&+ \left\langle \frac{dT_{m_1n_1}(p_1)}{dt}
T_{m_2n_2}(p_2) T_{m_3 n_3}(p_3) \right\rangle + (2 \text{ permutations}).}
Using the Ward identity the first line is an integral over a product of two-point functions, which we already know to be independent of $t$. Again we need the $dT/dt$ term to be of order $(tc)^n c^2$. Now this can happen, but only if we separate it into two correlators and have no factors of $t$. Such a term can only come from having one single-trace operator in the double-trace contribution to $T_{mn}^{(1)}$  combining with $T_{m_2 n_2}$, and the other with $T_{m_3 n_3}$. 
So in this term we get a product of two correlators of $T$ with an operator that is independent of $t$, and the same argument we used above implies that these $2$-point functions are independent of $t$.
Thus, we find that the $3$-point functions are linear in $t$, and their large $t$ behavior (for large $c$ at fixed $tc$) is determined just by the $2$-point functions in the original QFT before the deformation. We will see an explicit example of this in the next section. Note that since $t$ is dimensionful, taking large $t$ is the same as going to short distances or high energies (but still of order $\sqrt{|t|c}$).

Similarly, if we consider the $t$-derivative of a connected $4$-point function, we obtain on the first line on the right-hand side a product of a $3$-point function and a $2$-point function, which we already know to be linear in $t$. On the second line we can get a contribution from the double-trace term in $T_{mn}^{(1)}$, which gives a product of a $3$-point function (which can be of order $t$) with a $2$-point function (that is independent of $t$). Or, we can get a contribution from the triple-trace term in $T_{mn}^{(2)}$, which gives $t$ times a product of three $2$-point functions (which are independent of $t$). All other contributions are suppressed. So $4$-point functions are exactly quadratic in $t$ in this limit.
Similarly we find that, to all orders in perturbation theory in $t$, and to leading order in $1/c$, connected $n$-point functions are polynomials of degree $(n-2)$ in $t$, and they scale as $t^{n-2} c^{n-1}$ at large $t$. Moreover, they are simply related to the lower-point functions in the original QFT before the deformation; the leading large $t$ behavior is in fact determined just by $2$-point functions in the original QFT.
A similar behavior arises also for correlation functions of other ``single-trace operators''.

A large class of large $c$ theories has a weakly coupled (and sometimes also weakly curved) gravitational dual, by gauge/gravity duality (for CFTs this is given by a gravitational theory on $AdS_3$). In this language, the energy-momentum tensor is dual to a graviton in the bulk, and the deformation \eqref{tt} is given (in the large $c$ limit with fixed $tc$), like other ``double-trace deformations'' \cite{Aharony:2001pa,Witten:2001ua,Berkooz:2002ug}, by a change in the boundary condition for the graviton (see \cite{Marolf:2006nd,Compere:2008us}). On general grounds it is clear that perturbatively computing correlation functions in the gravitational theory precisely reproduces our analysis above, and this is consistent with the explicit computation of some correlation functions in \cite{Kraus:2018xrn}. If one tries to go away from the large $c$ limit, the deformation on the gravitational side is more drastic and the behavior near the boundary is significantly modified; it was suggested in \cite{McGough:2016lol} that this corresponds to putting a finite cutoff in the gravitational theory, but there are many ways of putting such a cutoff and it is not clear which, if any, of them corresponds to the correct deformation. Since $1/c$ corrections correspond to loop corrections on the gravity side, it is difficult to perform comparisons of the two sides away from the large $c$ limit.

\section{Example : $3$-point functions in the deformed CFT of a free scalar}

Using the methods described above, we can compute the 
three-point functions of energy-momentum tensors in a conformal field theory deformed by $T\Tb$, in the large $c$ limit. We use the notations $T=T_{++}$, $\Tb=T_{--}$ and $\Theta=T_{+-}$ for the different components of the energy-momentum tensor, where in Euclidean signature $x_{\pm}=x_1\pm i x_2$.
Following equation \eqref{dif}, the deformation will depend only on the two-point functions; however we need to keep also the contact terms in the two-point functions, which have non-trivial effects since the deformation is integrated. In momentum-space, the two-point functions including contact terms (which are determined by conservation) are given by 
\begin{center}
\begin{tabular}{ l l l }
$\langle T(p)T(-p)\rangle=-\frac{c}{12}\frac{p_+^3}{p_-}$, &$\langle T(p)\Tb(-p)\rangle=-\frac{c}{12}p_+p_-$, & $\langle T(p)\Theta(-p)\rangle=\frac{c}{12}p_+^2$, \\ 
 $\langle \Tb(p)\Tb(-p)\rangle=-\frac{c}{12}\frac{p_-^3}{p_+}$, &$\langle \Tb(p)\Theta(-p)\rangle=\frac{c}{12}p_-^2$, &$\langle \Theta(p)\Theta(-p)\rangle=-\frac{c}{12}p_+p_-$.
  
\end{tabular}
\end{center}

There are a total of ten different three-point correlation functions. We will do a detailed calculation for $\langle T(p)\Tb(q)\Theta(-p-q)\rangle$ and then state the results for the rest. Following \eqref{dif} we get
\eql{ex}{\frac{d}{dt}&\langle T(p)\Tb(q)\Theta(-p-q)\rangle\\=&256\int d^2w\Big[\frac{1}{w_-}\langle \Theta(-p-q)\Theta(-w)\rangle\left(w_+\langle T(p)\Tb(q)\Tb(w)\rangle+w_-\langle T(p)\Tb(q)\Theta(w)\rangle\right)\\&+\frac{1}{w_+}\langle \Theta(-p-q)\Theta(-w)\rangle\left(w_-\langle T(p)\Tb(q)T(w)\rangle+w_+\langle T(p)\Tb(q)\Theta(w)\rangle\right)\\&+\frac{1}{w_-}\langle \Tb(q)\Theta(-w)\rangle\left(w_+\langle T(p)\Theta(-p-q)\Tb(w)\rangle+w_-\langle T(p)\Theta(-p-q)\Theta(w)\rangle\right)\\&+\frac{1}{w_+}\langle \Tb(q)\Theta(-w)\rangle\left(w_-\langle T(p)\Theta(-p-q)T(w)\rangle+w_+\langle T(p)\Theta(-p-q)\Theta(w)\rangle\right)\\&+\frac{1}{w_-}\langle T(p)\Theta(-w)\rangle\left(w_+\langle \Tb(q)\Theta(-p-q)\Tb(w)\rangle+w_-\langle \Tb(q)\Theta(-p-q)\Theta(w)\rangle\right)\\&+\frac{1}{w_+}\langle T(p)\Theta(-w)\rangle\left(w_-\langle \Tb(q)\Theta(-p-q)T(w)\rangle+w_+\langle \Tb(q)\Theta(-p-q)\Theta(w)\rangle\right)\Big]\\&+\langle \frac{dT(p)}{dt}\Tb(q)d\Theta(-p-q)\rangle+\langle T(p)\frac{d\Tb(q)}{dt}d\Theta(-p-q)\rangle+\langle T(p)\Tb(q)\frac{d\Theta(-p-q)}{dt}\rangle.}

For the first contributions we can apply the Ward identities \eqref{ward}, which for this case take the form (see, for instance, \cite{Osborn:1993cr,Bzowski:2017poo})
\eq{p^{n_1}\langle &T_{m_1n_1}(p)T_{m_2n_2}(q)T_{m_3n_3}(-p-q)\rangle\\&=2p_{(m_3}\langle T_{n_3)m_1}(q)T_{m_2n_2}(-q)\rangle+2p_{(m_2}\langle T_{n_2)m_1}(-p-q)T_{m_3n_3}(p+q)\rangle\\&+(p+q)_{m_1}\langle T_{m_2n_2}(q)T_{m_3n_3}(-q)\rangle-q_{m_1}\langle T_{m_2n_2}(-p-q)T_{m_3n_3}(p+q)\rangle,}
with the brackets denoting symmetrization with a coefficient $\frac{1}{2}$. 
The contributions from this part are universal, independent of the specific theory, and we find
\eq{\frac{d}{dt}\langle  T(p)\Tb(q)\Theta(-p-q)\rangle=-\frac{16c^2}{9}\frac{p_+}{p_-}\frac{q_-}{q_+}(p_+q_--p_-q_+)^2+\text{non-universal terms},}
with the non-universal terms coming from the explicit derivatives $dT_{mn}/dt$.

Similarly, for the other three-point functions, we get
\eq{
 &\frac{d}{dt}\langle T(p)T(q)T(-p-q)\rangle=-\frac{8c^2}{9}\frac{p_++q_+}{p_-+q_-}\frac{p_+}{p_-}\frac{q_+}{q_-}(p_+q_--p_-q_+)^2 +\text{non-universal terms},\\ &\frac{d}{dt}\langle T(p)T(q)\Theta(-p-q)\rangle=\frac{-16c^2}{9}\frac{p_+}{p_-}\frac{q_+}{q_-}(p_+q_--p_-q_+)^2 +\text{non-universal terms},\\ 
 &\frac{d}{dt}\langle T(p)\Theta(q)\Theta(-p-q)\rangle=\frac{40c^2}{9}\frac{p_+}{p_-}(p_+q_--p_-q_+)^2+\text{non-universal terms}, \\  
 &\frac{d}{dt}\langle \Theta(p)\Theta(q)\Theta(-p-q)\rangle=-\frac{64c^2}{9}(p_+q_--p_-q_+)^2+\text{non-universal terms},\\
 &\frac{d}{dt}\langle T(p)T(q)\Tb(-p-q)\rangle=-\frac{8c^2}{9}\frac{p_+}{p_-}\frac{q_+}{q_-}\frac{p_-+q_-}{p_++q_+}(p_+q_--p_-q_+)^2+\text{non-universal terms},\\
 &\frac{d}{dt}\langle \Tb(p)\Tb(q)T(-p-q)\rangle=-\frac{8c^2}{9}\frac{p_-}{p_+}\frac{q_-}{q_+}\frac{p_++q_+}{p_-+q_-}(p_+q_--p_-q_+)^2+\text{non-universal terms},\\&\frac{d}{dt}\langle \Tb(p)\Tb(q)\Tb(-p-q)\rangle=-\frac{8c^2}{9}\frac{p_-}{p_+}\frac{q_-}{q_+}\frac{p_-+q_-}{p_++q_+}(p_+q_--p_-q_+)^2+\text{non-universal terms},\\&\frac{d}{dt}\langle \Tb(p)\Tb(q)\Theta(-p-q)\rangle=\frac{-16c^2}{9}\frac{p_-}{p_+}\frac{q_-}{q_+}(p_+q_--p_-q_+)^2+\text{non-universal terms},\\&\frac{d}{dt}\langle \Tb(p)\Theta(q)\Theta(-p-q)\rangle=\frac{40c^2}{9}\frac{p_-}{p_+}(p_-q_+-p_+q_-)^2+\text{non-universal terms}.}
Many of these results are pure contact terms, or contact terms of two of the operators, but we still write them since they will be important for going to higher-point correlation functions.

In a given theory there will be specific expressions for $dT_{mn}/dt$, and, as described above, any ``double-trace'' terms that this includes will give a contribution to the large $c$ $3$-point functions, which is a product of $2$-point functions of the $t=0$ theory. These contributions are essential for the Ward identities to continue to hold after the deformation.
In the case of the free scalar, $dT_{mn}/dt$ was explicitly computed in \cite{Cavaglia:2016oda}, and it is easy to generalize this to the case of $N$ free scalars, with $c=N$.
In that case each $dT_{mn}/dt$ is a ``double-trace operator''. The extra contributions to each correlation function are a sum of three products of two $2$-point functions, but the only non-contact term contribution in this case comes from $d\Theta/dt \propto T \Tb$.
%


%

Summing up the two types of contributions for the free scalar theory, the $3$-point functions that acquire a $t$-dependence that is not a contact term
are, in position space\footnote{These results agree with those of \cite{Kraus:2018xrn}, even though the details of the computation are different.}
\eql{noncontact}{\frac{d}{dt}&\langle  T(x)\Tb(y)\Theta(0)\rangle=32\frac{c^2}{x_+^4y_-^4}
,
\\\frac{d}{dt}&\langle \Tb(x)\Tb(y)T(0)\rangle=\frac{128c^2}{3}\bigg(\frac{1}{(y_--x_-)^5x_+^3}+\frac{1}{(x_--y_-)^5y_+^3}
\bigg),
\\\frac{d}{dt}&\langle T(x)T(y)\Tb(0)\rangle=\frac{128c^2}{3}\bigg(\frac{1}{(y_+-x_+)^5x_-^3}+\frac{1}{(x_+-y_+)^5y_-^3}
\bigg).
}
So the 3-point functions at separated points at finite $t$ will include $\langle TTT \rangle$ and $\langle \Tb \Tb \Tb \rangle$ that take the same value as in the CFT (proportional to $c$), and the additional ones coming from \eqref{noncontact} that are proportional to $tc^2$. At short distances (compared to $\sqrt{|t|c}$) the latter will dominate. Our results show no sign of any non-locality in these theories, even though the distances involved can be much smaller than the Hagedorn scale. In particular, they are compatible with still having an operator product expansion for the local operators, that includes new terms going as
\eql{ope}
{T(x) \Theta(0) & \sim \frac{t c \Tb(0)}{x_+^4}, \qquad \Tb(x) \Theta(0) \sim \frac{t c T(0)}{x_-^4}, \qquad \Tb(x) T(0) \sim t c \left(\frac{\partial_- \Tb(0)}{x_+^3} - \frac{\partial_+ T(0)}{x_-^3} \right), \\
T(x) T(0) & \sim \frac{t c x_- \Tb(0)}{x_+^5}, \qquad \Tb(x) \Tb(0) \sim \frac{t c x_+ T(0)}{x_-^5}.}

\section*{Acknowledgements}

We would like to thank S. Dubovsky, E. Gerchkovitz, A. Giveon, N. Itzhaki, O. Mamroud, V. Narovlansky, A. Schwimmer, G. Torrents, R. Yacoby, S. Yankielowicz, A. B. Zamolodchikov, and especially D. Kutasov, for many useful discussions. 
This work was supported in part  by the I-CORE program of the Planning and Budgeting Committee and the Israel Science Foundation (grant number 1937/12), by an Israel Science Foundation center for excellence grant, and by the Minerva foundation with funding from the Federal German Ministry for Education and Research. OA is the Samuel Sebba Professorial Chair of Pure and Applied Physics. TV is supported by the ERC STG grant 335182.

\bibliographystyle{JHEP}
\bibliography{TTbib}
\end{document}